\documentclass[aps,prd,amssymb,groupedaddress,nofootinbib]{revtex4}

\usepackage{epsfig}
\usepackage[english]{babel}

\usepackage{amsfonts}
\usepackage{amssymb}
\usepackage{amsmath}
\usepackage{slashbox}
\usepackage{multirow}

\def\d{\delta}

\def\={\nonumber &=}

\def\&{{}&}

\def\({\left(}
\def\){\right)}
\def\[{\left[}
\def\]{\right]}
\def\<{\left\langle}
\def\>{\right\rangle}

\def\bn{{\bf n}}

\def\curl{\mathcal}

\def\eq{\begin{align}}
\def\qe{\end{align}}
\def\eqa{\begin{eqnarray}}
\def\qea{\end{eqnarray}}

\def\and{\quad \mbox{and} \quad}

\def\bfnl{\kern2pt\overline{\kern-2ptf}_\textrm{NL}}

\def\lmax{l_\textrm{max}}

\def\aone{a_{l_1 m_1}}
\def\atwo{a_{l_2 m_2}}
\def\athree{a_{l_3 m_3}}
\def\afour{a_{l_4 m_4}}

\def\barQ{\kern2pt\overline{\kern-2pt\curl{Q}}}

\def\barR{\kern2pt\overline{\kern-2pt\curl{R}}}

\def\nmax{n_\textrm{max}}

\def\gnl{g_{\rm NL}}
\def\tnl{t_{\rm NL}}
\def\Gnl{G_{\rm NL}}

\def\setsize{\csname @setfontsize\endcsname \setsize}

\begin{document}


\title{Optimal Trispectrum Estimators and WMAP Constraints}

\author{J.R.~Fergusson}

\author{D.M.~Regan}

\author{E.P.S.~Shellard}

\affiliation
{Centre for Theoretical Cosmology,\\
Department of Applied Mathematics and Theoretical Physics,\\
University of Cambridge,
Wilberforce Road, Cambridge CB3 0WA, United Kingdom}

\date{\today}

\begin{abstract}
We present an implementation of an optimal CMB trispectrum estimator which accounts for 
anisotropic noise and incomplete sky coverage.   We use a general separable mode expansion which 
can and has been applied to constrain both primordial and late-time models.   We validate our methods on 
large angular scales using known analytic results in the Sachs-Wolfe limit.  We present the first
near-optimal trispectrum constraints from WMAP data on the cubic term of  local model inflation $
g_{\rm NL} = (1.6 \pm 7.0)\times 10^5$,  
for the equilateral model $t_{\rm NL}^{\rm{equil}}=(-3.11\pm 7.5)\times 10^6 $ and for the constant model 
$t_{\rm NL}^{\rm{const}}=(-1.33\pm 3.62)$.   These results, particularly the equilateral constraint,
are relevant to a number of well-motivated models (such as DBI and K-inflation) with closely correlated
trispectrum shapes.  We also use the trispectrum signal predicted for cosmic strings to provide a conservative
upper limit on the string tension $G\mu \le 1.1\times 10^{-6}$ (at 95\% confidence), which is
largely background and model independent.   All these new trispectrum results are consistent 
with a Gaussian Universe. We discuss the importance of constraining general classes of trispectra
using these methods and the prospects for higher precision with the Planck satellite.
\end{abstract}


\maketitle

\setsize{10}{12}

\section{Introduction}
Measurements of the cosmic microwave background (CMB) and large scale structure, such as those provided by the WMAP satellite or the Sloan Digital Sky Survey (SDSS), agree well with the predictions of standard single field slow-roll inflation. In particular the power spectrum verifies the prediction of a nearly scale invariant spectrum of adiabatic perturbations with a Gaussian distribution. However, there remains the prospect that significant non-Gaussianities may be produced by well-motivatived cosmological models which are consistent with the measured power spectrum. In order to quantify this non-Gaussianity we need to measure higher order correlators, beyond the two-point function or power spectrum.   In \cite{Regan:cn2010}, we presented an optimal estimation methodology to obtain the four-point correlator or trispectrum.  This was developed from the general bispectrum estimator 
of  \cite{FLS09,FLS10} which used separable mode expansions to investigate a much wider class of models than previously had been
investigated, as well as to reconstruct the full CMB bispectrum of the Universe.   These general bispectrum results were consistent with a Gaussian 
distribution.  However, it is possible for the three-point correlator to remain small but for there to be a large four-point correlator (see, for example, some 
inflationary models  \cite{aChen} and cosmic strings \cite{Regan}).  It is our purpose here to continue to test the standard inflationary 
paradigm through an optimised and expanded search for a trispectrum signal in the WMAP data. 

\par
In this paper we consider the class of four-dimensional trispectra which are independent of the diagonal. Such models include the cubic term of the local model, the equilateral model and the so-called `constant' model.   However, as we shall discuss, a general search for non-diagonal trispectrum shapes encompasses a much wider class of models, including most trispectra currently discussed in the literature. We demonstrate this for the non-diagonal equilateral shape which exhibits a high degree of correlation with the (apparently) higher dimensional shapes predicted by DBI and K 
inflation.   Another example is the cosmic string trispectrum which can be reduced to a closely correlated non-diagonal shape.  Such dimensional reduction is an important first step in testing even truly diagonal shapes because trispectrum estimation in the non-diagonal case is much more straightforward. Exploiting the use of a separable expansion, as we do here, ensures the 
computation is tractable, reducing the complexity from $\mathcal{O}(l_{\rm{max}}^7)$ to $\mathcal{O}(l_{\rm{max}}^4)$, and it also ensures the stability of the algorithms used in the analysis without the need to correct for pathological terms commonly present in other approaches. 

\par
The constraints obtained here on the cubic term of the local model, the equilateral model and the `constant' model  result from comparison to year 5 WMAP data out to $l=500$ together with a pseudo-optimal analysis of inhomogeneous noise and masking contributions. The estimators and other algorithms employed here were outlined in detail in \cite{Regan:cn2010}, but are expressed in this paper with the simplifying assumption that the trispectrum is independent of the diagonal term.  We validate our results by using known analytic results in the large angle limit \cite{RSF10} where the signal-to-noise can be calculated explicitly.  This is important because we are able to show that previous trispectrum forecasts using this Sachs-Wolfe approximation were over-optimistic (see, for example, the interesting analysis of trispectrum forecasts in refs~\cite{Kogo,DS}).   We note that we will describe the implementation of these methods and the reconstruction of the CMB trispectrum in much more detail in a longer accompanying paper \cite{FRS11}. 

\par 
Some alternative approaches to extracting information about the primordial trispectrum from the WMAP data have been explored in the literature. In \cite{VIELVA} the constraint $-5.6\times 10^5<g_{NL}<6.4\times 10^5$ was obtained by analysing the $N$-point probability distribution of CMB anisotropies (a non-optimally-weighted method, see \cite{Regan:cn2010}).  This work assumed a local perturbative model, $\Phi=\Phi_L+f_{NL}(\Phi_L^2+\langle \Phi_L^2\rangle)+g_{NL}\Phi_L^3$, while using a Sachs-Wolfe approximation out to  $\theta < 1^\circ$.    However, we note 
that our calculation of the {\it optimal} WMAP variance at $\lmax \approx 500$ has error bars of $\pm 10.7 \times 10^5$ (at 95\% confidence).   
 In ref.~\cite{Cooray,Cooray2}, a local trispectrum constraint  $-7.4 \times 10^5 < g_{\rm NL} < 8.2 \times 10^5$  is estimated from WMAP using a kurtosis power spectrum analysis (along with a constraint on $\tau_{\rm NL}$).   However, this approach does not directly subtract the effect of anisotropic noise and other systematic effects using the quadratic terms in the optimal trispectrum estimator \cite{Regan:cn2010}; we know from the present work that these are important in obtaining an accurate and optimized result.   We will discuss and compare these alternative approaches in more detail in our longer paper.  It is also appropriate at this stage to mention other earlier proposals for measuring the four-point correlator using a harmonic analysis on COBE\cite{0111250} and a wavelet approach \cite{0301220}.

\par
n section II we review results relating primordial and CMB trispectra and their optimal estimation, here, focusing on primordial (and hence CMB) trispectra which are independent of the diagonal.  The eigenmode decomposition of the trispectrum constitutes the basis of our method and is reviewed in section III.  In section IV we obtain constraints on the cubic local model term, the equilateral model and the constant model. The accuracy of this approach is also verified, using large-angle analytic results for the local $\gnl$ model. We emphasise the close correlation of our non-diagonal equilateral model with 
a broader set of equilateral-type models. We demonstrate the application of this method also at late times, when we constrain the cosmic 
string trispectrum and discuss forecasts for the Planck data.  Finally we discuss our results and present our conclusions in section V.  

\section{Optimal Trispectrum Estimation}

CMB temperature anisotropies
may be represented using the $a_{lm}$ coefficients of a spherical harmonic decomposition of the cosmic microwave sky,
\begin{eqnarray}\label{DeltaT}
\frac{\Delta T}{T}(\hat{\mathbf{n}})=\sum_{l,m} a_{lm}Y_{lm}(\hat{\mathbf{n}}).
\end{eqnarray}
The primordial potential $\Phi$ induces the multipoles $a_{lm}$ via a convolution with the transfer functions $\Delta_l(k)$,
\begin{eqnarray}
a_{lm}=4\pi (-i)^l \int \frac{d^3k}{(2\pi)^3} \Delta_l(k) \Phi(\mathbf{k}) Y_{lm}(\hat{\mathbf{k}}).
\end{eqnarray}
The connected part of the four-point correlator of the $a_{lm}$ gives us the trispectrum, 
\begin{eqnarray}\label{Tconnls}
T_{l_1 m_1 l_2 m_2 l_3 m_3 l_4 m_4}&=&\langle a_{l_1 m_1}	 a_{l_2 m_2} a_{l_3 m_3} a_{l_4 m_4}	\rangle_c \nonumber\\
&=&(4\pi)^4 (-i)^{\sum_i l_i}\int  \frac{d^3 k_1 d^3 k_2 d^3 k_3 d^3 k_4}{(2\pi)^{12}} \;\Delta_{l_1}(k_1)\Delta_{l_2}(k_2)\Delta_{l_3}(k_3)\Delta_{l_4}(k_4)~\times~\cr
&& \langle \Phi(\mathbf{k_1})\Phi(\mathbf{k_2})\Phi(\mathbf{k_3})\Phi(\mathbf{k_4})\rangle_c  \;Y_{l_1 m_1}(\hat{\mathbf{k_1}}) Y_{l_2 m_2}(\hat{\mathbf{k_2}}) Y_{l_3 m_3}(\hat{\mathbf{k_3}}) Y_{l_4 m_4}(\hat{\mathbf{k_4}}),
\end{eqnarray}
where $k_i=|\mathbf{k}_i|$ and the subscript $c$ denotes the connected component. In this paper we will specialise to statistically isotropic trispectra which depend only on the wavenumbers $k_1,k_2,k_3,k_4$, that is, the `diagonal-free' trispectra class (and the many models which are closely correlated 
with these).  
Using this assumption we may write 
\begin{eqnarray}\label{eq:Tconn}
\langle \Phi(\mathbf{k_1})\Phi(\mathbf{k_2})\Phi(\mathbf{k_3})\Phi(\mathbf{k_4}) \rangle_c=(2\pi)^3\, \delta (\mathbf{k_1+k_2+k_3+k_4})\;T_{\Phi}(k_1,k_2,k_3,k_4).
\end{eqnarray}
Substituting this into \eqref{Tconnls}, it is straightforward to show that 
\begin{eqnarray}\label{TconnlsNew}
T_{l_1 m_1 l_2 m_2 l_3 m_3 l_4 m_4}&=& \int d\Omega_{\hat{\bn}}\;Y_{l_1 m_1}(\hat{\bn})Y_{l_2 m_2}(\hat{\bn})Y_{l_3 m_3}(\hat{\bn})Y_{l_4 m_4}(\hat{\bn})\; t^{l_1 l_2}_{l_3 l_4},
\end{eqnarray}
where the `extra'-reduced trispectrum $t^{l_1 l_2}_{l_3 l_4}$ is given by (see \cite{RSF10})
\begin{align}\label{eq:extraRed}
t^{l_1 l_2}_{l_3 l_4}=&\left(\frac{2}{\pi}\right)^4 \int x^2 dx \int (k_1 k_2 k_3 k_4)^{2}\;T_{\Phi}(k_1,k_2,k_3,k_4)\;\Delta_{l_1}(k_1)\Delta_{l_2}(k_2)\Delta_{l_3}(k_3)\Delta_{l_4}(k_4)\nonumber\\
&\qquad\qquad\qquad\qquad\times j_{l_1}(k_1 x) j_{l_2}(k_2 x) j_{l_3}(k_3 x) j_{l_4}(k_4 x).
\end{align}
As with the bispectrum analysis it is simpler to analyse the trispectrum - especially for scale-invariant models - in terms of a shape function, i.e. a scale-invariant version of the trispectrum. Removing a $k^{-9}$ scaling in the diagonal-free case would motivate the following shape function
\begin{align}\label{eq:Shape}
S(k_1,k_2,k_3,k_4)=(\Delta_{\Phi}^3 N)^{-1}{(k_1 k_2 k_3 k_4)^{9/4}}\;T_{\Phi}(k_1,k_2,k_3,k_4)
\end{align}
where $\Delta_{\Phi}$ gives the amplitude of the Sachs-Wolfe peak of the power spectrum and $N$ is a normalisation factor which may be identified with $9t_{NL}/200$ such that  $S(k,k,k,k)=1$ (see \cite{aChen}).
Naively, the numerical calculation of the CMB trispectrum  (\ref{eq:extraRed}) appears extremely challenging (even in this diagonal-free case), as it involves a line of sight integral over a highly oscillatory 4D integral. However, if the shape can be represented in separable form $S(k_1,k_2,k_3,k_4)=W(k_1)X(k_2)Y(k_3)Z(k_4)$, then   the integral breaks down into a much simpler product of one-dimensional integrals.

In order to compare theoretical predictions with observation it is necessary to use an estimator which sums the signal-to-noise over the range of multipoles probed, essentially performing a least squares fit of the data to the theoretical trispectrum $\langle \aone^{\rm{th}} \atwo^{\rm{th}} \athree^{\rm{th}} \afour^{\rm{th}} \rangle_c$. In \cite{Regan:cn2010}, the general optimal trispectrum estimator was derived giving
\begin{eqnarray}\label{eq:Estimator2}
\mathcal{E}&=&\frac{f_{\rm{sky}}}{\tilde{N}^2}\sum_{l_i m_i} \langle a_{l_1m_1}a_{l_2m_2}a_{l_3m_3}a_{l_4m_4}\rangle_c  \Bigg[(C^{-1} a^{\rm{obs}})_{l_1 m_1}(C^{-1} a^{\rm{obs}})_{l_2 m_2} (C^{-1} a^{\rm{obs}})_{l_3 m_3} (C^{-1} a^{\rm{obs}})_{l_4 m_4}\nonumber\\
&&\qquad\qquad \qquad -6(C^{-1})_{l_1 m_1,l_2 m_2} (C^{-1} a^{\rm{obs}})_{l_3 m_3} (C^{-1} a^{\rm{obs}})_{l_4 m_4}+3 (C^{-1})_{l_1 m_1,l_2 m_2}(C^{-1})_{l_3 m_3,l_4 m_4} \Bigg],
\end{eqnarray}
where $f_{\rm{sky}}$ is the sky fraction observed,  the covariance matrix $C_{l_1 m_1,l_2 m_2}\equiv\langle \aone \atwo\rangle$ is non-diagonal due to mode-mode coupling introduced by the mask and anisotropic noise, and $N$ is the appropriate normalization  $N$ (see ref.~\cite{Regan:cn2010} for further details).  Issues of optimality for the bispectrum estimator corresponding to \eqref{Estimator2} were addressed in ref.~\cite{0503375} and
further relevant references are reviewed in \cite{ReviewLFSS}.  We note that a related expression for optimal estimation of the trispectrum in large-scale structure (and other higher-order correlators)  was presented in ref.~\cite{FRS10}.

In this paper we follow \cite{07114933,WMAP5} by assuming a nearly diagonal covariance matrix ($C_{l_1 m_1,l_2 m_2}\approx (-1)^{m_1} C_{l_1}\d_{l_1 l_2}\d_{m_1 -m_2}$) and account for the noise $N_l$ and instrument beam $b_l$ by setting 
$C_l\rightarrow\tilde{C}_{l}=b_l^2 C_l+N_l$ and ${t}^{l_1 l_2}_{l_3 l_4}\rightarrow\tilde{t}^{l_1 l_2}_{l_3 l_4}=b_{l_1}b_{l_2}b_{l_3 }b_{l_4}{t}^{l_1 l_2}_{l_3 l_4} $ (we drop tildes henceforth).   
With this identification the estimator becomes
\begin{eqnarray}\label{eq:estim}
\mathcal{E}=\frac{f_{\rm{sky}}}{N_T^2}\sum_{l_i m_i}  \frac{\langle a_{l_1m_1}a_{l_2m_2}a_{l_3m_3}a_{l_4m_4}\rangle_c }{\tilde{C}_{l_1} \tilde{C}_{l_2} \tilde{C}_{l_3} \tilde{C}_{l_4}}\Big[a^{\rm{obs}}_{l_1 m_1}a^{\rm{obs}}_{l_2 m_2}a^{\rm{obs}}_{l_3 m_3}a^{\rm{obs}}_{l_4 m_4}-6\,C_{l_1 m_1, l_2 m_2}^{\rm{sim}}a^{\rm{obs}}_{l_3 m_3}a^{\rm{obs}}_{l_4 m_4}
+ 3\,C_{l_1 m_1, l_2 m_2}^{\rm{sim}}C_{l_3 m_3, l_4 m_4}^{\rm{sim}}\Big],
\end{eqnarray}
where $N_T^2$ for the non-diagonal trispectrum reduces to the weighted sum 
\begin{align}\label{eq:Normal}
N_T^2=&\sum_{l_i}w_{l_1l_2l_3l_4}\frac{{t}^{l_1 l_2}_{l_3 l_4}{t}^{l_1 l_2}_{l_3 l_4}}{\tilde{C}_{l_1}\tilde{C}_{l_2}\tilde{C}_{l_3}\tilde{C}_{l_4}}\quad \hbox{with }\quad w_{l_1l_2l_3l_4}= {\textstyle{ \frac{(2l_1+1)(2l_2+1)(2l_3+1)(2l_4+1)}{2 (4\pi)^2}}}\int_{-1}^1 P_{l_1}(\mu) P_{l_2}(\mu) P_{l_3}(\mu) P_{l_4}(\mu)d\mu\,.
\end{align}
The second and third terms in \eqref{eq:estim} ensures subtraction of spurious inhomogeneous noise and masking by using the covariance matrix $C_{l_1 m_1, l_2 m_2}^{\rm{sim}}$ from an ensemble average of Gaussian maps in which these effects are incorporated. Again, if the theoretical trispectrum $t^{l_1 l_2}_{l_3 l_4}$ has the property of primordial separability then the summations in \eqref{eq:estim} and \eqref{eq:Normal} become much more tractable, taking only $\mathcal{O}(l_{\rm{max}}^4)$ operations.  Finally, it is important to note that the autocorrelator given by \eqref{eq:Normal} can be used to define a natural measure on the allowed trispectrum domain and an inner product between bispectra
$ \langle {t},{t'}\rangle = \sum_{l_i}w_{l_1l_2l_3l_4}{{t}^{l_1 l_2}_{l_3 l_4}{t'}^{l_1 l_2}_{l_3 l_4}}/{{C}_{l_1}{C}_{l_2}{C}_{l_3}{C}_{l_4}}$.

It should be noted that is desirable to compare measures of the trispectra of different models. The general measure, $t_{NL}$, is defined in the equilateral limit for which the wavenumbers $k_i$ and the diagonals of the quadrilateral (formed by the wavevectors $\mathbf{k}_i$) have equal values, i.e. $k_1=k_2=k_3=k_4=|\mathbf{k}_1+\mathbf{k}_2|=|\mathbf{k}_1+\mathbf{k}_3|=k$. We define
\begin{align}\label{tnldef}
t_{NL}=\frac{9}{200}\frac{T_{\Phi}(k,k,k,k;k,k)}{P_{\Phi}(k)^3}.
\end{align}
With this definition we find that for the local model $t_{NL}^{\rm{local}}=1.5 \tau_{NL}+1.08 g_{NL}$ (see \cite{aChen}). 

\par
As with the the shortcomings of using the parameter $f_{NL}$ to compare different bispectra consistently (see \cite{FLS09}), the method of comparing trispectra using their value at a central point is problematic.
Choosing a different choice of normalisation factor to $N_T^2$ allows us to define an integrated measure with which a consistent comparison between models may be made. In \cite{RSF10} the following choice of normalisation factor was adopted $N^2\equiv N_{T}N_{T\rm{locB}}$, where $\rm{locB}$ refers to the $g_{NL}$ expression of the local model trispectrum (with $g_{NL}=1$ and $\tau_{NL}=0$). The $g_{NL}$ term is chosen since is independent of the diagonal term and therefore may be determined more quickly than the normalisation for the $\tau_{NL}$ term. Hence we adopt the following definition for a general meaure of the trispectrum
\begin{align} \label{eq:GNLdef}
\Gnl =\frac{N_T}{N_{T\,{\rm loc}}^{\gnl=1}}\mathcal{E}.
\end{align}
This choice removes the large disparities that can exist between quoted constraints for different models.  It is also applicable to
non-scale invariant models, as well as trispectra induced by late-time processes such as gravitatioal lensing and cosmic strings. An approximation scheme for evaluating $N_T$ using primordial shape correlators has been outlined in \cite{RSF10}.

\section{Separable Mode Expansion}
In \cite{Regan:cn2010} separable mode expansions were derived for a general (reduced) trispectrum. In this paper we consider trispectra which are 
independent of the diagonal, so the complexity in finding a general separable expansion reduces from $\mathcal{O}(l_{\rm{max}}^5)$ to $\mathcal{O}(l_{\rm{max}}^4)$.  Here we will consider separable mode expansions at both early times and late times in order 
to project forward an arbitrary non-diagonal primordial trispectrum $T(k_1,k_2,k_3,k_4)$ and convert it into a weighted expansion for the late-time
CMB trispectrum ${t}^{l_1 l_2}_{l_3 l_4}$.   
We make a separable mode expansion of the primordial shape function \eqref{eq:Shape} in the form
\begin{align}\label{eq:shapedecomp}
S(k_1,k_2,k_3,k_4)=\sum_n \alpha_n^{Q}\,Q_n(k_1,k_2,k_3,k_4),
\end{align}
where the $Q_n$ are product functions $Q_n(k_1,k_2,k_3,k_4)=q_{(p}(k_1) q_r(k_2) q_s(k_3) q_{t)} (k_4)$
and $n=\{p,r,s,t\}$, with $(p r s t)$ representing the $24$ cyclic permutations reflecting the underlying trispectrum symmetry.
As we have emphasised for bispectrum mode expansions \cite{FLS09}, the one-dimensional basis functions can be chosen for convenience provided 
they are well-behaved and nearly scale-invariant.   Here, we choose analogs of Legendre polynomials but with a weight function set by the 4D 
integration domain allowed by the quadrilateral condition, that is, 
$k_1\leq k_2+k_3+k_4$ for $k_1\geq k_2,k_3,k_4 + \rm{cyclic\,perms}$ (denoted ${\cal V}_T$).
We may integrate arbitrary functions $f(k_1,k_2,k_3,k_4)$ and $g(k_1,k_2,k_3,k_4)$ over this domain, 
defining an inner product $\langle f,g\rangle\equiv\int_{\mathcal{V}_T}f \, g \,\omega\, d\mathcal{V}_T$, where $\omega$ is a given weight function (here $\omega=1$).  

The $Q_n$ modes though independent and separable are not, in general, orthonormal, i.e. $\langle Q_n ,Q_p\rangle=\gamma_{np}\neq \delta_{np}$, so it is often more useful to work in terms of orthonormal basis functions  $R_n$ with $\langle R_n, R_p\rangle= \delta_{np}$ 
(created from the $Q_n$ via a Gram-Schmidt process). The relation between the $R_n$ and $Q_n$ modes is given by
\begin{align}
R_n=\sum_{p=0}^n \lambda_{n p}Q_p, \quad\textrm{where}\quad (\lambda_{np})^{-1}=\langle R_n  Q_p\rangle\,,
\end{align}
where the matrices $\gamma_{np}$ and $\lambda_{np}$ are related by
$\gamma_{np}^{-1}=\sum_r \lambda_{r n}\lambda_{r p}$.
Just as in (\ref{eq:shapedecomp}), the shape function may be expanded in terms of the orthonormal $R_n$ basis $
S(k_1,k_2,k_3,k_4)=\sum_n \alpha_n^{R}R_n(k_1,k_2,k_3,k_4)$, where the $\alpha_n^R=\langle R_n, S\rangle$ are related to the $\alpha_n^Q$ through $\alpha_n^Q=\sum_p \lambda_{p n}\alpha_p^{R}$.

\par
A similar decomposition may be applied to the late-time trispectrum $t^{l_1l_2}_{l_3 l_4}$ with separable product and orthonormal modes, denoted 
with a bar as $\overline{Q}_n(l_1,l_2,l_3,l_4)\equiv \overline{q}_{\{p}(l_1) \overline{q}_r(l_2) \overline{q}_s(l_2) \overline{q}_{t\}}(l_4)$ and $\overline{R}_n(l_1,l_2,l_3,l_4)$ respectively (with the $\overline {q}_p(l)$ defined on the allowed multiple domain and a chosen ordering $n\leftrightarrow \{prst\}$).   Here, we use the late-time inner product defined below \eqref{eq:Normal}
with $\overline{\gamma}_{n m} =\langle \overline{Q}_n, \overline{Q}_m\rangle$.  We expand the trispectrum  signal-to-noise in the estimator as
\begin{align}\label{eq:latetime}
\frac{v_{l_1}v_{l_2}v_{l_3}v_{l_4}}{\sqrt{C_{l_1}C_{l_2}C_{l_3}C_{l_4}}}\;t^{l_1l_2}_{l_3 l_4}=\sum_n \overline{\alpha}_n^Q \;\overline{Q}_n(l_1,l_2,l_3 l_4).
\end{align}
where, for the sake of scale invariance, we use the freedom to multiply by a separable weight function, here with $v_l= (2l+1)^{1/4}$.
Given the form of the (diagonal-free) primordial shape decomposition \eqref{eq:shapedecomp} we can project 
forward the $Q_n(k_1,k_2,k_3,k_4)$ modes to late times  and express these in terms of the 
$\overline{Q}_n(l_1,l_2,l_3 l_4)$ used in (\ref{eq:latetime}).   Explicitly, defining the convolved basis functions as $q^l_p(x) = \frac{2}{\pi}\int dk\,q_p(k)\, \Delta_l(k)j_l(kx)$ and using \eqref{eq:extraRed} and \eqref{eq:shapedecomp},  we obtain
\begin{align}\label{eq:earlytime}
\overline{\alpha}^Q_n = \sum_p\Gamma_{np} \, \alpha_p^Q\,\quad \hbox{with}\quad \Gamma_{np} = \int dx\, x^2\, \tilde{\gamma}_{nr}(x) \;\overline{\gamma}^{\,-1}_{rp}
\end{align}
where $\tilde{\gamma}_{n p}(x)= \langle\tilde{Q}_n^{l_1 l_2 l_3 l_4}(x),\, \overline{Q}_p(l_1,l_2,l_3,l_4)\rangle\,$,  $\tilde{Q}_n^{l_1 l_2 l_3 l_4}(x)=\tilde{q}_r^{l_1}(x)\tilde{q}_s^{l_2}(x) \tilde{q}_t^{l_3}(x) \tilde{q}_u^{l_4}(x)$ and $ \tilde{q}_t^{l}(x) =  v_l {q}_t^{l}(x)
/\sqrt{C_{l}}$  (see also \cite{FLS09}). 
The huge efficiency advantage of using the late-time decomposition \eqref{eq:latetime} is that the line-of-sight integration is captured once and for all 
in the transformation matrix $\Gamma_{np}$ and need not be repeated in any subsequent Fisher matrix and estimator analysis.

Substituting the diagonal-free mode expansion \eqref{eq:shapedecomp} into  \eqref{eq:estim}, and exploiting 
\eqref{eq:earlytime}, the estimator separates and reduces to the simple summation
\begin{align}\label{eq:modalest}
\mathcal{E}&~=~\frac{\Delta_{\Phi}^3 N}{N_T^2}\sum_n \overline{\alpha}^Q_n\; \overline{\beta}^Q_n~=~ \frac{\Delta_{\Phi}^3 N}{N_T^2}\sum_{n,p} \Gamma_{np}\,{\alpha}^Q_p\; \overline{\beta}^Q_n
\end{align}
where the observed $\beta^Q_n$ coefficients arise from separable products of the CMB map filtered with the polynomial basis functions 
$\overline {q}_p(l)$,
\begin{align}
M_{p}(\hat{\bf{n}})=\sum_{l m}\frac{q_p(l)a_{l m}Y_{l m}(\hat{\bf{n}})}{\sqrt{C_l}(2l+1)^{1/4}}\,.
\end{align}
Explicitly, there are quartic, quadratic and constant (unconnected) parts  $  \overline{\beta}^Q_n =  \overline{\beta}^{(4)}_n - 6 \overline{\beta}^{(2)}_n +  3 \overline{\beta}^{(0)}_n$ defined by 
\begin{align}\label{betadefs}
\beta^{(4)}_n \equiv \int d^2{\hat {\bf n}}\; M_{\{p}\,M_{r}\,M_{s}\,M_{t\}}\,,\quad \beta^{(2)}_n \equiv \int d^2{\hat {\bf n}}\; \langle M_{\{p}\,M_{r}\rangle 
\,M_{s}\,M_{t\}} ~~\hbox{and}~~\beta^{(0)}_n \equiv \int d^2{\hat {\bf n}}\; \langle M_{\{p}\,M_{r}\rangle \langle M_{s}\,M_{t\}}\rangle\,,
\end{align}
where the angled brackets represent map products averaged over many Monte Carlo simulations incorporating realistic anisotropic noise, beam and masking effects. Remarkably, once the mode coefficients and transformations have been calculated,  trispectrum estimation using (\ref{eq:modalest}) collapses down to only ${\cal O}(\lmax^2)$ operations.

\section{WMAP constraints on non-diagonal models}

We have applied the full estimator \eqref{eq:estim} using separable mode expansions \eqref{eq:modalest} to the WMAP data for a variety 
of non-diagonal models.   At present, this work represents the first near-optimal CMB trispectrum analysis, incorporating the subtraction 
of systematic anisotropic effects,  but it 
neither exhausts the full precision available in the WMAP data nor the broad array of models which can be constrained through 
this general analysis.  Here we report the first step in implementing this formalism in a CMB trispectrum analysis for examples of the 
simplest models (at both primordial and late times).  We note that we employ the pseudo-optimal
weighting in  \eqref{eq:estim}, rather than the  inverse covariance weighting in \eqref{eq:Estimator2}, analysing the WMAP 5-year dataset 
out to multipoles with $\lmax \le 500$.   We coadded the V and W band data, using the same weights, noise model, beams and KQ75 mask 
as in the WMAP5 analysis.   A similar modal bispectrum analysis was performed successfully in refs.~\cite{FLS09,FLS10} and we shall describe the extensive work undertaken with this trispectrum estimator more fully in our longer paper
\cite{FRS11}.   However, here we note that the extraction of $\overline{\beta}^Q_n$ coefficients \eqref{eq:modalest} from 400 Gaussian maps in a 
fully WMAP-realistic context produced no significant bias in our trispectrum estimator. Working at $\lmax =500$, we found it sufficient to use
only about fifty modes in our separable decompositions \eqref{eq:shapedecomp} and \eqref{eq:latetime} to adequately 
describe all the models under investigation  (i.e.\ $\nmax =50$).

\subsection{Local Model (cubic $g_{\rm NL}$ term)}

\begin{figure}[htp]
\centering 
\includegraphics[width=122mm]{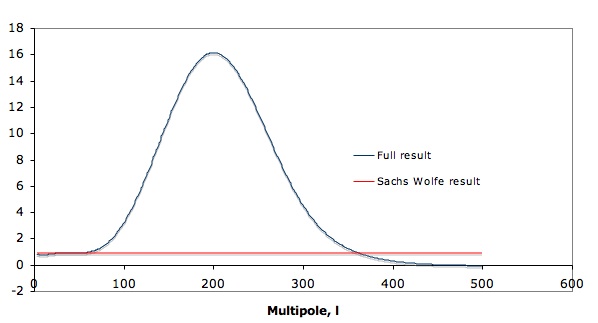}
\caption{Plot of the ratio of the full model decomposition trispectrum with the analytic Sachs-Wolfe result \eqref{eq:SachsWolfegnl} for the local $g_{NL}$ model. The excellent agreement for $l\lesssim 60$ demonstrates the accuracy of the modal formalism adopted in this paper. }
\label{fig:equalLtrisp}
\end{figure}

Local non-Gaussianity is one of the best-motivated models for non-Gaussianity because it can be 
produced by local interactions in minimal extensions of the standard model of inflation,
such as those involving multiple scalar fields (see, for example, the comprehensive review in \cite{Chen}). 
The $\gnl$ term of the local model arises as the coefficient of the cubic term in a Taylor expansion
around the linear fluctuations $\Phi_{\rm L}$:
\begin{align}
\Phi=\Phi_{\rm L}+f_{\rm NL}(\Phi_{\rm L}^2+\langle \Phi_{\rm L}^2\rangle)+g_{\rm NL}\Phi_{\rm L}^3+\mathcal{O}(\Phi_{\rm L}^4).
\end{align}
Note that there is also a local trispectrum contribution from the square of the quadratic term (in single field inflation 
this is simply $\tau_{\rm NL} = (6f_{\rm NL}/5^2$), but this is much difficult to analyse because it involves integration
over two solid angles (a treatment we shall discuss in \cite{FRS11}).    The cubic $\gnl$ term induces the following 
non-diagonal trispectrum shape (see \cite{Okamoto}):
\begin{align}\label{eq:shapelocalgnl}
S_T^{g_{NL}}(k_1,k_2,k_3,k_4)=\frac{(k_1 k_2 k_3 k_4)^{9/4}}{24 \Delta_{\Phi}^3}T_{g_{NL}}(k_1,k_2,k_3,k_4)=\frac{k_1^3+k_2^3 +k_3 ^3 +k_4^3}{4(k_1 k_2 k_3 k_4)^{3/4}}.
\end{align}
In the Sachs-Wolfe approximation where we take the simple transfer function $\Delta_l(k)=\frac{1}{3}j_l((\tau_0-\tau_{\mathrm{dec}})k)$, we obtain
the large-angle solution \cite{Regan:cn2010}
\begin{align}\label{eq:SachsWolfegnl}
t^{l_1 l_2}_{l_3 l_4}=\frac{2\Delta_{\Phi}^3 }{27 \pi^3}\left(\frac{1}{l_2(l_2+1)l_3(l_3+1)l_4(l_4+1)}	~+~3\;\mathrm{perms}\right).
\end{align}
This analytic solution provides an excellent benchmark against which to test the accuracy of our modal decomposition methodology.  
In  Figure~\ref{fig:equalLtrisp} we plot the equal-$l$ trispectrum $t^{\,l\, l}_{\,l\, l}$ for the analytic Sachs-Wolfe solution \eqref{eq:SachsWolfegnl} and 
the actual trispectrum calculated using the full transfer functions starting from the primordial decomposition coefficients $\alpha^Q_n$ in 
\eqref{eq:shapedecomp} through the transformation matrix $\Gamma_{np}$ in \eqref{eq:latetime}.   The agreement is excellent, despite the rather pathological local $\gnl$ shape with 
divergences on faces at each $k_i\rightarrow 0$.  We were able to obtain an accurate primordial decomposition 
(to within a 90\% correlation using a CMB-motivated cut-off as in \cite{Fergusson:2008ra}), 
while the $\Gamma_{np}$ conversion to the late time decomposition \eqref{eq:latetime} introduced insignificant further 
correlation errors (at around the 1\% level).    Both factors can be mitigated further by increasing the mode number $\nmax$.   

\par
It is instructive to note the behaviour of the trispectrum in Figure~\ref{fig:equalLtrisp} due to the extra power of the transfer function 
which enhances Silk damping suppression.  The rapid asymptotic fall-off  seen at $l\approx 400$ is actually preceded by 
a decline in the signal to noise (relative to the $C_l$'s) which starts at $l\ge 60$ in the first acoustic peak.  Making projections
 beyond $l\ge 60$ using the Sachs-Wolfe signal-to-noise estimate leads to over-optimistic expectations \cite{Kogo,DS}, also
indicating that the results in \cite{vielva2,Cooray2} are somewhat super-optimal. 
The optimal (Fisher) bound at 1$\sigma$ significance is given by 
\begin{align}
\Delta {g_{NL\,\mathrm{opt}}}=\frac{1}{(S/N)}=\frac{\sqrt{24 }}{\sqrt{f_{sky}}N_T^{g_{NL}}}=5.35\times 10^5.
\end{align}

\par
For our WMAP5 $\gnl$ analysis we extracted the observed trispectrum coefficients $\overline{\beta}^Q_n$ using the estimator \eqref{eq:modalest}. 
This entailed subtraction of unconnected and systematic anisotropic contributions using the expectation of the filtered maps in \eqref{betadefs} 
averaged over very many Gaussian realisations (see longer paper \cite{FRS11}).   We note, here, that the effect of the anisotropic noise and mask made
contributions at least at the level of the optimal variance, necessitating its careful subtraction (achieved here for the first time).  A direct comparison 
of the observed orthonormal   $\overline{\beta}^R_n$ and the predicted orthornormal $\gnl$ coefficients $\overline{\alpha}^R_n$ demonstrated no 
significant correlation between modes.    Substitution into the modal estimator \eqref{eq:modalest} yielded the constraint 
\begin{align}
\gnl \equiv \Gnl =(1.62\pm 6.98)\times 10^5\,.
\end{align}
The variance here was obtained from $400$ Gaussian WMAP5 realisations, noting that we obtain a variance approximately 30\% above 
the optimal bound at this resolution (in line with a previous modal bispectrum analysis \cite{FLS10}).   In order to achieve a result closer to the optimal bound the same modal methodology can be used but after pre-conditioning the CMB data with inverse covariance matrix weighting, as in \eqref{eq:Estimator2}.   We conclude that there appears to be no evidence for a local $\gnl$
non-Gaussian signal in the WMAP data.

\subsection{Constant Model}

The constant model represents the simplest possible primordial shape with 
\begin{align}
S_T^{\rm{const}}(k_1,k_2,k_3,k_4)=\frac{(k_1 k_2 k_3 k_4)^{9/4}}{24 \Delta_{\Phi}^3}T_{\rm{const}}(k_1,k_2,k_3,k_4)=1\,.
\end{align}
This shape yields a CMB trispectrum $t^{l_1 l_2}_{l_3 l_4}$ with features entirely due to the transfer functions.   As for the bispectrum, it provides a useful benchmark for identifying features which can be expected to be shared by all primordial models. 
The trivial primordial decomposition with only $\alpha^Q_0\ne 0$ is 100\% accurate, but it still achieves above 99\% accuracy for the late-time decomposition \eqref{eq:latetime} after the $\Gamma_{np}$ transformation using  $50$ modes.  The optimal (Fisher) bound for the constant model is given by 
\begin{align}
\Delta {t_{NL\, \mathrm{opt}}^{\mathrm{const}}}=\frac{1.08}{(S/N)}=1.08\frac{\sqrt{24 }}{\sqrt{f_{sky}}N_T^{\mathrm{const}}}=2.67\times 10^6.
\end{align}
Again a direct comparison of the observed WMAP spectrum $\overline{\beta}^R_n$ and the predicted constant model $\overline{\alpha}^R_n$  
indicated no obvious correlation.  Of some interest was a negative zeroth WMAP mode $\overline{\beta}^R_0$ about which we shall 
comment later.    The modal estimator \eqref{eq:modalest} yielded the (equivalent) constant model constraints
\begin{align}
\tnl^{\rm{const}}=(-1.33\pm 3.62)\times 10^6\quad\Longleftrightarrow \quad \Gnl =(-2.64\pm 7.20 )\times 10^5.
\end{align}
where the $t_{NL}$ parameter was defined in \eqref{tnldef} and the universally normalized trispectrum parameter $\Gnl$ was defined in \eqref{eq:GNLdef}.   The estimated variance was again determined from $400$ Gaussian simulations in the same WMAP-realistic context. We conclude that there is no evidence from the WMAP data for a constant primordial trispectrum signal.  

\subsection{Equilateral Model}

Equilateral-type models are produced through the amplification of nonlinear effects around the time the modes exit the horizon. A non-standard kinetic term allows for such a possibility.
In \cite{aChen} it was shown that the leading order trispectrum for single field inflation models is described by a combination of three scalar-exchange trispectra, $T_{s_1,s_2,s_3}$, and three contact-interaction trispectra, $T_{c_1,c_2,c_3}$ (detailed formulae for these models can be found in ref.~\cite{aChen}). These trispectra generally have a diagonal dependence, however, there is one diagonal-free model $T_{c_1}$ which we shall denote the (canonical) {\it equilateral model} with a shape function defined by
\begin{align}\label{eq:shapeequil}
S_T^{c_1}(k_1,k_2,k_3,k_4)=\frac{(k_1 k_2 k_3 k_4)^{9/4}}{24 \Delta_{\Phi}^3}T_{c_1}(k_1,k_2,k_3,k_4)=\frac{(k_1 k_2 k_3 k_4)^{5/4}}{\left((k_1+k_2+k_3+k_4)/4\right)^{5}}.
\end{align}
Therefore, we shall consider the correlation of this non-diagonal equilateral model $T_{c_1}$ with those that have some diagonal dependence.  
This requires a full 5D Monte Carlo integration to determine the equilateral cross-correlations yielding (see \cite{FRS11} for further details):

\par
\begin{center}
\begin{tabular}{|c|c|c|c|c|c|}

 \hline
 
(Model A, Model B) &         $(c_1,s_1)$&   $(c_1,s_2)$ & $(c_1,s_3)$  &$(c_1,c_2)$ & $(c_1,c_3)$ \\

    \hline
 $\mathcal{C}$(Model A, Model B)  &$0.95$&$0.88$&$0.82$ &$0.70$ &$0.56$\\
    
    \hline
  \end{tabular}
\end{center}

\noindent In particular, we note that the equilateral model correlation with the $K$-inflation prediction  \cite{PDM99,WY08,ELL08} 
($T_k\propto T_{s_3}$)
is thus 82\%, while DBI inflation  \cite{AlishahihaSilversteinTong2004, aChen} which is a linear combination of $T_{c_1}$ and $T_{s_1,s_2,s_3}$  exhibits an
83\% correlation.  Indeed, almost all single field inflation models of interest are well approximated by considering the equilateral $T_{c_1}$ model. 
Hence, we draw the important conclusion that any limit on the non-diagonal equilateral model will also strongly constrain other well-motivated 
equilateral-type models. 

The primordial eigenmode expansion \eqref{eq:shapedecomp} with $50$ modes correlates with the shape \eqref{eq:shapeequil} 
at about $\sim 99.7\%$ accuracy. We then project this primordial shape forward to make a CMB prediction with the late-time decomposition in  \eqref{eq:earlytime}. 
A Fisher matrix analysis reveals an optimal variance for the equilateral model of 
\begin{align}
\Delta {t_{NL\,\mathrm{opt}}^{\mathrm{equil}}}=5.52\times 10^6.
\end{align}
A direct comparison between the equilateral model and recovered mode coefficients for the WMAP5 data reveals little correlation, just as with the
constant model. The modal estimator \eqref{eq:modalest} produces the (equivalent) equilateral constraints
\begin{align}
t_{NL}^{\rm{equil}}=(-3.11\pm 7.5)\times 10^6 \quad\Longleftrightarrow \quad G_{NL}=(-3.02\pm 7.27)\times 10^5.
\end{align}
We conclude that WMAP provides no evidence in favour of a wide range of equilateral-type models, including $K$-inflation, 
$DBI$ inflation and all the other shapes in the table above  identified above with single field inflation.

\subsection{Cosmic strings}

A significant advantage of the formalism developed here (as in \cite{RSF10, FLS09}) is that it may be readily applied to late-time models as well as primordial models. Here, we apply the methodology to the trispectrum of cosmic strings. Cosmic strings are line-like discontinuities which may be formed during a phase transition in the very early universe \cite{KibbleMech} or at the end of brane inflation (in which case they are often denoted 
cosmic superstrings \cite{TyeSarangi}).  Cosmic strings are characterised by their tension $G\mu$. Our intention, therefore, is to use the trispectrum induced by cosmic strings to constrain $G\mu$.
\par
In \cite{Regan:2009hv} the trispectrum induced by cosmic strings was derived in a WMAP and Planck context (see also \cite{hind10}). The analysis assumed that the temperature discontinuity produced by cosmic strings is given entirely by the Gott-Kaiser-Stebbins effect \cite{Gott,KAISERSTEBBINS}. The quantity derived originally  \cite{Regan:2009hv} was inclusive of the unconnected term and included a diagonal dependence. However, an 
accurate approximation, which correlates with the full trispectrum within $\sim 90\%$ is given by
\begin{align}\label{eq:pcombi}
(l_1 l_2 l_3 l_4)^{3/2}p^{l_1 l_2}_{l_3 l_4}=&(8\pi G\mu)^4\frac{2\overline{v}^4 \pi }{s^2 }\frac{ l_2^2}{(l_1 l_2 l_3 l_4)^{1/2}}   \frac{1}{1.63} \frac{l_1^2\tilde{\xi}^2}{(0.63+l_1\tilde{\xi})} \ln\left(\frac{1+\eta_0/\eta_{lss}}{2}\right)\left( \frac{2}{1+500/l_m}\right)^{2.3},
\end{align}
where we note that $l_m=\min(500,l_i)$, $\tilde{\xi}=1/l_m$ and $\eta_0/\eta_{lss}\approx 50$. The values of the parameters $\overline{v}^2$ and $s^2$ are found by simulations \cite{MartShell,BBS,AllShel} to be given by the numerical values $\overline{v}^2=0.365$ and $s^2=0.42$. 
The full trispectrum is then given by the permutations 
$t^{l_1 l_2}_{l_3 l_4}=p^{l_1 l_2}_{l_3 l_4}+p^{l_1 l_3}_{l_2 l_4}+p^{l_1 l_4}_{l_2 l_3}$  (for further details see \cite{Regan2011}).
We then decompose the cosmic string trispectrum in the form \eqref{eq:latetime},  obtaining an expansion which is  found to have a 
 $99\%$ correlation at 50 modes. The signal to noise $S/N=N_T\sqrt{f_{sky}/24}$ at $\lmax <500$ gives the following optimal bound on the cosmic string tension 
achievable using the CMB trispectrum,
\begin{align}
G\mu \lesssim 8.8\times 10^{-7}.
\end{align}
This should be compared to current CMB power spectrum constraints on Abelian-Higgs strings $7\times 10^{-7}$ \cite{Bevis} and on Nambu-Goto strings $2.5\times 10^{-7}$ \cite{Battye}. We have also obtained forecasts for the optimal tension that may be probed using multipoles up to $l_{\rm{max}}=2000$ (achievable using Planck data). We find for $l_{\rm{max}}=1000,1500$ and $2000$ that the optimal tension probed is given respectively by $G\mu_{\rm{opt}}=4.3\times 10^{-7},\; 2.8\times 10^{-7}$ and $1.8\times 10^{-7}$.  Unlike the local $g_{NL}$ model the signal to noise for cosmic strings is unaffected by Silk damping. Thus, as a test for cosmic strings, we expect the trispectrum to provide a competitive probe to the power spectrum given the increased resolution of the Planck satellite and it may prove to be the most stringent probe in future surveys.
\par
We can compare the cosmic string trispectrum decomposition from \eqref{eq:pcombi} to that obtained from the WMAP$5$ data.   The modal estimator~\eqref{eq:modalest}, yields the $1\sigma$ bound
\begin{align}
\left(\frac{G\mu}{2\times 10^{-7}}\right)^4= -852\pm 870.
\end{align}
where the variance is again obtained from $400$ Gaussian simulations.   Since the string tension must be positive, we deduce from this
anti-correlation that cosmic strings are disfavoured by current CMB data. The $2\sigma$ bound on cosmic strings is then given by $1.1\times 10^{-6}$.
However, we note the caveat that the cosmic string spectrum is dominated by a positive constant term $\overline{\alpha}^R_0>0$, which is observed in 
the WMAP data to be negative $\overline{\beta}^R_0<0$.   As discussed previously for the bispectrum \cite{FLS09}, the constant term $\overline{\beta}^R_0$ is susceptible to contamination by  poorly resolved point sources, a possibility which we will continue to investigate.

\section{Discussion}
We have presented results from an implementation of an optimal CMB trispectrum estimator which employs separable mode expansion applicable to a wide class of isotropic models that are diagonal independent, i.e.\ that only depend on the wavenumbers $k_1,k_2,k_3,k_4$. We have obtained constraints on the cubic term for the local model, the constant model and a notable new constraint on equilateral-type models. We found no evidence for primordial non-Gaussianity for these trispectrum shapes (or models closely correlated with them) at the 95\% confidence level. 
The constraints on the parameter $g_{NL}^{local}$ represent a constraint on the self-interaction term of the local model and, in addition, allows for a qualitative classification of models of the local type \cite{1009.1979}. The constraints presented here on the equilateral model are entirely new results. The importance of finding bounds on such models is, not least, as a consequence of the high correlation between the equilateral model, $K$-inflation and DBI inflation. 

\par
In addition we have demonstrated that this estimator methodology can be applied to late-time models by searching for the characteristic
trispectrum shape induced by cosmic strings.   We have also obtained a conservative bound on cosmic strings $G\mu\lesssim 1.1\times 10^{-6} $(at $95\%$ confidence). Using forecasts for the signal to noise at Planck resolution we have established that the trispectrum of cosmic strings is expected to give comparable constraints to the power spectrum in the near future.  This constraint has the advantage of not exhibiting degeneracies with background 
cosmological parameters, while also being largely independent of the underlying properties of the cosmic strings (unlike gravitational wave constraints).

\par 
This paper marks a significant first step in a general analysis of trispectrum models. Among the advantages of the modal approach pursued here are 
the exploitation of efficiencies arising from separability and the transfer function transformation, the absence of pathologies when representing 
shapes, and the opportunity to investigate non-separable models which were previously deemed intractable.   We also note  that this general 
mode expansion allows for an accurate characterisation of the noise and foregrounds which must be subtracted to achieve an optimal estimation measure.  While the constraints published in this paper are consistent with Gaussianity, the extension of the analysis to general trispectra represents an important development \cite{FRS11}. Such an analysis will, for instance, allow a measurement of the local trispectrum parameter $\tau_{NL}$ providing a  test of local inflation which requires $\tau_{NL}\geq(6f_{NL}/5)^2$.  An implementation of the late-time modal estimator outlined in \cite{RSF10} to 
include trispectra dependent on the diagonal term will allow identification of any trispectrum whether generated at primordial times like inflation or late-times like gravitational lensing or second-order gravitational effects. This, in conjunction with recent results classifying CMB bispectrum constraints
\cite{FLS10}, offers the hope of a comprehensive test for non-Gaussianity. 

\par 
We conclude by noting that we have obtained near-optimal trispectrum constraints on primordial local $\gnl$ and equilateral (and correlated) models, 
together with late-time constraints on cosmic strings.    We find no evidence for a significant trispectrum in these non-diagonal cases, about  
which we will provide greater detail elsewhere \cite{FRS11}, together with a reconstruction of the CMB trispectrum obtained from the WMAP data.   
The new results here represent a significant test of the simplest standard model of inflation, further 
affirming the Gaussian hypothesis for primordial fluctuations.

\section*{Acknowledgements}
We are very grateful for many informative and illuminating discussions with 
Xingang Chen and with Michele Liguori, with whom the modal bispectrum estimator was
developed. EPS and JRF were supported by STFC grant ST/F002998/1 and the 
Centre for Theoretical Cosmology.  DMR was supported by EPSRC and the Isaac Newton Trust.
Simulations in this paper were performed on the COSMOS National Cosmology Supercomputer 
supported by STFC, DBIS, SGI and Intel.  We are very grateful for expert computational help from 
Andrey Kaliazin.

\bibliography{TrispConstraint}

\end{document}